\documentstyle[preprint,version2,aps]{revtex}
\draft
\begin{document}

\begin{title}
Critical Properties of the Mott Transition in the Hubbard Model
\end{title}

\author{Goetz Moeller, Qimiao Si,\\
Gabriel Kotliar and Marcelo Rozenberg}

\begin{instit}
Serin Physics Laboratory, Rutgers University,
Piscataway, NJ 08855-0849, USA
\end{instit}

\begin{abstract}

We introduce a systematic low-energy approach to strongly correlated
electron systems in infinite dimensions, and apply it to the problem
of the correlation-induced metal-insulator transition in the
half-filled Hubbard model. We determine the low-energy scaling
functions of the metallic state, including the single-particle Green
function and  dynamical spin susceptibility, as well as  thermodynamic
properties  and relate them to experimental data in transition
metal oxides.
\end{abstract}
\pacs{PACS numbers: 71.27+a, 74.20Mn, 71.28+d, 71.10+x}

\newpage
%\begin{narrowtext}

The correlation driven metal-insulator transition is a fundamental
problem in condensed matter physics which has attracted renewed
theoretical and experimental interest. From a theoretical point of
view, the problem is fascinating because of its non-perturbative
nature, the absence of a small parameter and the emergence of new low
energy scales. Early ideas of Mott\cite{Mott}, Hubbard\cite{Hubbard},
Brinkman and  Rice\cite{Brinkman} have recently
been put on a more quantitative footing by the development of a
mean-field approach to the strong correlation problem which becomes
exact in the limit of large lattice coordination\cite{Vollhardt}. The
work of several groups\cite{Jarre,Marcelo1,GeoKra1} has established
the existence of a Mott transition in the paramagnetic phase of the
half-filled Hubbard model on frustrated  lattices, which is driven by the
collapse of an energy scale,
the renormalized Fermi energy. The presence of this small energy scale
made it difficult to analyze the critical behavior near the transition.

In this letter, we introduce a novel projective self-consistent approach
to solving correlated electron problems in large dimensions at low
energies. We use the natural
separation of energy scales to our advantage  and are able to extract,
for the first time, exact information about the critical behavior at
the  transition in infinite dimensions.
We calculate the single particle spectral function near the Mott
transition, which can in principle be measured in photoemission
experiments. We establish  that the coefficient of the
$\omega^2$ term in the imaginary part of the self-energy diverges as
the square of the coefficient  of the linear term in the specific
heat and, with few additional assumptions, relate it to the observed
$T^2$ resistivity in the $LaSrTiO_3$ system. We show that the local
spin  susceptibility diverges in the same way as the linear term in
the specific heat
and calculate their ratio, a generalized Wilson number.
Finally, we show that the linear coefficient of the imaginary part of
the dynamical spin susceptibility, relevant to both NMR and neutron
scattering experiments, diverges as the square of the linear
coefficient of the specific heat.

Besides these results of theoretical and experimental interest, our
method can be generalized to deal with other problems, in which
the separation of energy scales typical of strongly correlated
electron systems can be exploited\cite{eps}.
Two examples are the breakdown of Fermi liquid theory
in strongly correlated electron systems\cite{SK} and superconductivity
in models with repulsive interactions\cite{GeoKot3,Werner3}, which
have been recently analyzed using the large-d techniques.

The Hamiltonian of the Hubbard model is given by
\begin{eqnarray}
{\rm H} = -\sum_{<ij> \sigma} (t_{ij} +\mu \delta_{ij}) f^{+}_{i\sigma}
f_{j\sigma}
 + U \sum_{i}
(f_{i\uparrow}^{\dagger}f_{i\uparrow} -{1 \over 2})
(f_{i\downarrow}^{\dagger}f_{i\downarrow} -{1 \over 2})
\label{hamiltonian}
\end{eqnarray}
where the chemical potential $\mu$ is taken to be zero, and the
hopping matrix $t_{ij}$ is scaled as $t_{ij} \rightarrow t/\sqrt{d}$.
In the limit of infinite dimensions, all  local correlation
functions of the lattice model can be calculated in terms of a local
action,
\begin{eqnarray}
{\rm S}_{\rm imp} [G_o]&=& -\int^{\beta}_0 d\tau d\tau '
\sum_{\sigma} f^{\dagger}_{\sigma}(\tau )G^{-1}_o(\tau-\tau')
f_{\sigma} ( \tau' )
 \nonumber\\ &&
+\int^{\beta}_0 d\tau {U}
(n_{f\uparrow}(\tau) -{1\over 2}) (n_{f\downarrow}(\tau)-{1 \over 2})
\label{siteaction}
\end{eqnarray}
provided that $G_o$ satisfies a self-consistency
equation\cite{GeoKot}. In order to make contact with physical systems
in finite dimensions, for which the bandwidth is finite, we consider the
case of a semi-circular bare density of states, $\rho_o
(\epsilon) = (2 / {\pi D}) \sqrt{1 - ( \epsilon /D)^2}$ for which the
self-consistency equation has the form
\begin{eqnarray}
G_o^{-1} (i \omega_n )= i\omega_n + \mu -({D / 2})^2G (i \omega_n ).
\label{selfcons}
\end{eqnarray}

The local action (\ref{siteaction}) can be rewritten in terms of an
Anderson impurity model\cite{GeoKot}
\begin{eqnarray}
{\rm H}_{\rm AM} =
\sum_{k \sigma} \epsilon_k
c_{k,\sigma}^{\dagger} c_{k,\sigma}  + \sum_{k\sigma} V_k (
f_{\sigma}^{\dagger} c_{k\sigma} +h.c.)
+U(n_{f\uparrow} -{1 \over 2})(n_{f\downarrow}-{1 \over 2}).
\label{anderson}
\end{eqnarray}
The self-consistency equation (\ref{selfcons}) is satisfied provided
that the bath dispersion $\epsilon_k$ and the hybridization coupling
$V_k$ are determined through
\begin{eqnarray}
\sum_k {4 V_k^2 / D^2 \over {i\omega_n -\epsilon_k}}= G(i\omega_n).
\label{consistency}
\end{eqnarray}
where $G(i\omega_n)=-\int_0^\beta d\tau e^{i \omega_n \tau}
<T_{\tau}f_{\sigma} (\tau ) f^{\dagger}_{\sigma}(0)>_{\rm H_{\rm
AM}}$ is the impurity Green function.

For sufficiently strong interactions, a separation of energy scales
occurs  as demonstrated explicitly by previous
numerical work on the Hubbard model\cite{Jarre,Marcelo1,GeoKra1}.
It has been shown that the one particle spectral function can be
decomposed into a sum of a  low and a  high energy part, $\rho
(\epsilon) =\rho^{low} (\epsilon )+\rho^{high} (\epsilon )$.
$\rho^{low}(\epsilon)$ contains all states up to a cut-off that we
take to be the Kondo temperature or renormalized Fermi energy of the
half-filled Hubbard model and carries spectral weight $\Delta$
\begin{equation}
\Delta = {\sum_{ k }}' {4V^2_{k}}/{D^2}
\end{equation}
where the primed summation runs over the low energy states only.
$\rho^{high} (\epsilon )$ describes the upper and lower Hubbard bands,
two atomic-like features at energy scales $\pm {U \over 2}$, and
carries spectral weight $1 -\Delta$. When $\Delta $ is small the
presence of the low energy scale $\Delta D$ makes the system of
equations (\ref{anderson}) and (\ref{consistency}) numerically
untractable.

The main idea of the projective self-consistent approach
is to eliminate the high energy states to obtain
a low energy  effective problem involving $\rho^{low}$ only and thus
containing  only one energy scale.
To carry out this program we separate the impurity configurations of
the impurity Hamiltonian (\ref{anderson}) into a low energy sector
$f_{\sigma}^{\dagger}|E>$, with eigenenergy  $-U/4$, and  a
high energy sector $|E>$ and
$f_{\uparrow}^{\dagger}f_{\downarrow}^{\dagger}|E>$, with
eigenenergy $U/4$, where $|E>$ denotes the empty configuration.
Furthermore, the self-consistency equation allows us to divide
the conduction electron bath into
three bands as illustrated in Fig. (\ref{bath}): a metallic band
centered around the Fermi energy, and semiconducing  valence  and
conduction bands centered around energies $\pm U/2$. For the
analysis of low energy
properties, the valence and conduction bands
can be taken as dispersionless\cite{note2}, with the corresponding
atomic states created by $\eta_{-}^{\dagger}$ and $\eta_{+}^{\dagger}$,
respectively.

The low energy effective Hamiltonian can be derived through a
canonical transformation. In addition to the high
energy impurity configurations, which can be eliminated by the standard
Schrieffer-Wolff\cite{SchWol} transformation, the self-consistency
requires the elimination of the valence and conduction bands of the
electron bath. For this purpose we first separate out the atomic Hamiltonian
\begin{eqnarray}
{\rm H}_{\rm atomic } = &&U(n_{f\uparrow} -{1 \over
2})(n_{f\downarrow}-{1 \over 2}) \nonumber\\
&&+ {U \over 2}
(\eta_{+,\sigma}^{\dagger} \eta_{+,\sigma}
- \eta_{-,\sigma}^{\dagger}\eta_{-,\sigma})
+V \sum_{l=+,-}\sum_{\sigma} (f_{\sigma}^{\dagger} \eta_{l\sigma}
+h.c.)
\end{eqnarray}
where $V$ is determined from the fact that both the upper and lower
Hubbard bands have a spectral weight
$(1-\Delta)$, which together with the self-consistency equation
(\ref{consistency}) implies that $V =(D/2)\sqrt{(1-\Delta)/2}$.
The lowest-energy eigenstates of this atomic Hamiltonian are a spin
doublet $|\sigma>$ which correspond to the impurity spin doublet
$f_{\sigma}^{\dagger}|E>$ dressed by the valence and conduction
electrons. All other atomic configurations are located at energies
higher by at least $\sim {U\over 2}$.

We have carried out the canonical transformation analytically to order
$(D/U)^2$\cite{note2,Kutz}, resulting in a low-energy effective
Hamiltonian ${\rm H}_{eff}= e^S {\rm H} e^{-S}$ of the form
\begin{equation}
{\rm H}_{eff} = {\sum_{k k'}}' J_{kk'} \vec S \cdot \vec s_{kk'} +
{\sum_{k \sigma}}' \epsilon_k c_{k \sigma}^{\dagger} c_{k \sigma}
\label{Kondo}
\end{equation}
where $J_{k k'} = 8 {V_k V_{k'} \over U} (1+{7\over 4}{D^2\over U^2}-
{7\over 4}{D^2\over U^2}\Delta)$, $\vec S= {1 \over 2} \sum_{\sigma
\sigma'} X_{\sigma,\sigma'} \vec \sigma_{\sigma \sigma'}$ and
$\vec s_{kk'}= {1\over 2} \sum_{\sigma \sigma'} c_{k,\sigma}^{\dagger}
\vec \sigma_{\sigma \sigma'} c_{k',\sigma'}$.
Here, $X_{\alpha \beta} \equiv |\alpha><\beta|$,
where  $|\alpha>$ and $|\beta>$ are the eigenstates of
the atomic Hamiltonian, are projection operators.

${\rm H}_{eff}$ corresponds to a Kondo problem coupling the atomic
spin doublet to the low energy conduction electron bath.
The low energy part of the Green  function can now be calculated
directly from $H_{eff}$,
\begin{equation}
G^{low}(i\omega_n) =
-\int d\tau e^{i\omega_n\tau} < T_{\tau} F(\tau) F^{\dagger}(0 ) >
_{{\rm H}_{eff}}\label{lowG}
\label{glow}
\end{equation}
where $F_{\sigma}={\rm e}^Sf_{\sigma}{\rm e}^{-S}$
is the canonically transformed single particle operator and has the form
\begin{equation}
F_{\sigma} = {\sum_{k}}' \alpha_k
[(X_{\sigma \sigma}- X_{-\sigma -\sigma} )c_{k \sigma}
+ 2 X_{\sigma -\sigma} c_{k -\sigma}]
\end{equation}
with $\alpha_k=\frac{2V_k}{U}
(1+{7\over 4}{D^2\over U^2}- {7\over 4}{D^2\over U^2}\Delta)$.
The self-consistency equation for the low energy Green function then
becomes
\begin{equation}
{\sum_{k}}' \frac{4V^2_k/D^2} {i\omega_n -\epsilon_k} = G^{low}(i\omega_n).
\label{sc.kondo}
\end{equation}
The projective self-consistent method thus results in
the closed set of equations (\ref{Kondo}-\ref{sc.kondo}) which
form the basis of our low energy analysis.
The system contains only
{\it one} energy scale, $\Delta D$, which allows us to
define rescaled variables,
$\tilde V_k = V_k /(\sqrt{\Delta} D)$, $\tilde \epsilon_k = \epsilon_k /
(\Delta D)$, $i \tilde \omega_n = i \omega_n / (\Delta D)$,
and $\tilde {\rm H}_{eff} = {\rm H}_{eff}
/(\Delta D)$.

In the following, we will focus on the critical properties of the Mott
transition approached from the metallic side, for which it suffices to
analyze Eqs. (\ref{Kondo}-\ref{sc.kondo}) to leading order in
$\Delta$. Restricting to leading order in $\Delta$ implies that
the requirement that the rescaled spectral weight is unity, $\int
d{\tilde \omega} \rho^{low}(\tilde \omega) =1$, can be satisfied
together with the self-consistency equations
(\ref{lowG}-\ref{sc.kondo}) only for one particular value of $U$. This
is precisely the critical interaction  $U_c$.

The problem is now reduced to a self-consistent determination of the
low-energy parameters $\tilde \epsilon_k$ and $\tilde V_k$, which is
carried out iteratively using a recently introduced zero-temperature
algorithm\cite{SiMar,KraCaf}. The functional equations
(\ref{Kondo}) and (\ref{sc.kondo}) are truncated by representing the
electron bath in terms of a {\it finite} set of free electron orbitals
$c_{i}^{\dagger}$, $i=1,...,N-1$, with the corresponding energy levels
$\epsilon_i$ and hybridization matrix $V_{i}$. The ground state of the
impurity problem, now defined on an $N$ site cluster, is obtained  by
the modified Lanczos technique\cite{Elb}. The local Green function as
well as correlation functions are calculated using a continued
fraction expansion\cite{Bals,SiMar}. The projected
self-consistency condition
(\ref{sc.kondo}) can be implemented in terms of a $\chi^2$ fitting of
the Matsubara Green function
${\chi^2 = \sum_{\tilde \omega_n=\Omega_{min}}^{\Omega_{max}} |\sum_k
\frac{\tilde V_k^2}{i \tilde \omega_n- \tilde \epsilon_k}-
 <F^{\dagger} F >_{\rm \tilde H_{\rm eff}} (i \tilde \omega_n) |^2}$
where $\Omega_{min}$ and $\Omega_{max}$ are the low and high frequency
cut-offs, respectively\cite{KraCaf}.

We have solved the equations for clusters of 4,6,8 and 10 sites.
The critical interaction $U_c$ is found to be essentially unchanged as
the number of sites is increased from 4 to 10, varying from
$3.03$ to $3.04$\cite{note}.
In the inset of Fig. (\ref{ImG}) we show the single particle spectral
function for $N=10$. The large number of poles clearly demonstrates the
power of the projection onto the low energy sectors.
The Matsubara Green functions as functions of the rescaled Matsubara
frequency for $N=6$, $8$ and $10$ are shown in Fig. (\ref{ImG}).  The
low frequency part of the Green function improves systematically as
$N$ increases, and remains essentially unchanged for $N=8$ and $10$.
Also shown in Fig. (\ref{ImG}) is the Green function for the non-interacting
problem. The value of the scaled Green function at zero frequency is
determined by the low energy self-consistency equation and is
the same as that of the Green function of the non-interacting problem,
consistent with the pinning of the density of states expected from
Fermi liquid theory\cite{Mueller-Hartmann}.
The solution for the Green function also allows us to determine the
scaled self-energy  $\Sigma^{low}(i \tilde \omega_n)$ by extracting
from
$\Sigma (i\omega_n ) =i \omega_n-(D/2)^2 G(i\omega_n) -
G^{-1} (i\omega_n)$ the terms with a singular dependence on $\Delta$.
Both $G^{low}(i\tilde \omega_n)$ and
$\Sigma^{low}  (i\tilde
\omega_n)$ can be fitted by quadratic polynomials at low frequencies
with the expressions

\begin{eqnarray}
G^{low}(i\tilde \omega_n) &&= -i 2.0 +3.3 (i\tilde \omega_n)+i 3.9 (i \tilde
\omega_n)^2 ,\nonumber\\
\Sigma^{low}(i\tilde \omega_n) &&= - 1.7 i\tilde \omega_n- i 1.1  ( i \tilde
\omega_n)^2.
\label{fit.selfe}
\end{eqnarray}

The term of the self-energy linear in $i\tilde \omega_n$  implies
a quasi-particle residue $z \equiv (1-\partial \Sigma/\partial
i \omega_n)^{-1} = \Delta  / 1.7$ which vanishes as the critical point
is approached. The momentum-independence of the self-energy in turn
leads  to a quasi-particle mass $m^*/m = 1/z= 1.7/\Delta$,  and a
linear coefficient of the specific heat,

\begin{eqnarray}
\gamma = {4\pi k_B^2 \over 3}{1.7 \over D \Delta}
\label{gamma}
\end{eqnarray}
which diverge at the critical point. This divergence is consistent with
the Brinkman-Rice scenario of the Mott transition\cite{Brinkman},  as
well as the previous result in the Hubbard model at half-filling
within second order perturbation
theory\cite{Marcelo2,GeoKra2}. The term of the
self-energy quadratic in $(i\tilde \omega_n)^2 $ gives rise to an
imaginary part of the analytically-continued self-energy

\begin{eqnarray}
Im \Sigma (\omega + i0^+) =-{1.1 \over \Delta ^2} {\omega^2 \over D}
\label{im.sigma}
\end{eqnarray}
which also diverges at the critical point.

The  local dynamical  spin susceptibility
$\chi_s (i \omega_n) = (g\mu_B/2)^2
\int_0^\beta d\tau {\rm e}^{i\omega_n \tau} <T_{\tau} S_z (\tau) S_z
(0)>_{H_{eff}},$
where $S_z = {1 \over 2} {\rm e}^S (f^{\dagger}_{\uparrow}f_{\uparrow}
-f^{\dagger}_{\downarrow}f_{\downarrow}) {\rm e}^{-S} = (X_{\uparrow
\uparrow} -X_{\downarrow \downarrow} )/2$, can also be calculated from
a continued fraction expansion. The result in terms of rescaled
frequency, $\chi_s(i\tilde \omega_n)$, is shown in Fig.(\ref{spin}).
At low frequencies, $\chi_s(i\tilde \omega_n)$ can be fitted by

\begin{eqnarray}
\chi_s^{low} (i\tilde \omega_n) = (g\mu_B/2)^2 (9-47 |\tilde
\omega_n|)/\Delta D
\label{fit.chi}
\end{eqnarray}

The local static susceptibility is given by the constant term of Eq.
(\ref{fit.chi}),

\begin{eqnarray}
\chi_s = (g\mu_B/2)^2 9/(\Delta D )
\label{chi0}
\end{eqnarray}
from which we derive a generalized Wilson ratio at the critical point,

\begin{equation}
R \equiv { \chi_s/\chi^{free}_{s,loc} \over \gamma/ \gamma^{free}}= 3.2
\end{equation}
where $\chi^{free}_{s,loc}=16/(3 \pi D)({g\mu_B \over 2})^2 $ and
$\gamma_{free}={4 \pi \over 3 D} k_B^2  $, are the  static local
spin susceptibility and the linear
coefficient of the specific heat for the free electron gas with a
semicircular density of states, respectively.
Compared to the universal value for the infinite bandwidth Anderson
impurity model,  $R_{AM}=2$, the critical value $R$ is enhanced
as  a result of  the finite bandwidth of the electron bath.
%We note that the usual definition of the Wilson ratio in a lattice
%model is given in terms of the ${\bf q}=0$ component of the static
%susceptibility.
%In large dimensions the ${\bf q}$ dependent susceptibility has a
%typical value $\chi_{loc}$ for a general ${\bf q}$ in the Brillouin
%zone which is different from the ${\bf q}=0$ value.
%,controlled by the magnetic exchange. In order to avoid the special
%features of the ${\bf q=0}$ point in large dimensions we have used
%$\chi_{loc}$ in the definition of the  generalized Wilson ratio. The
%enhancement of $R$ is consistent with the ferromagnetic tendencies
%found in the Gutzwiller approximation\cite{VollHe3} where the
%magnetic exchange is  ignored as well.\cite{footnote}
In large dimensions the susceptibility at a generic ${\bf q}$
is not affected by the antiferromagnetic exchange
interaction and, hence, has the typical value $\chi_{loc}$.
The enhancement of $R$ is consistent with the ferromagnetic tendencies
found in the Gutzwiller approximation\cite{VollHe3} where the magnetic
exchange is  ignored as well.
We note that the usual definition of the Wilson ratio in a lattice
model is given in terms of the ${\bf q}=0$ component of the static
susceptibility, $\chi_s ({\bf q=0})$. The antiferromagnetic exchange
interaction does affect $\chi_s ({\bf q=0})$ (as well as $\chi_s ({\bf
q})$ for special ${\bf q}$ points in the Brillouin zone), leading to a
vanishing ratio $\chi_s ( {\bf q=0} ) /\gamma$ at the transition
point.\cite{footnote}
%In order to avoid the special features of the ${\bf q=0}$ point in
%large dimensions we have used $\chi_{loc}$ in the definition of the
%generalized Wilson ratio.

Finally, the term of the imaginary part of the dynamical spin
susceptibility  linear in $|\tilde \omega_n|$ implies that
$\lim_{\omega \rightarrow 0} {\chi_s'' (\omega + i0^+ ) \over \omega}
= 47/(D \Delta)^2 (g \mu_B/2)^2$.
If we use $\chi_{loc}$ for the static spin susceptibility this implies
a finite generalized Korringa ratio at the critical point, which is
again modified from the universal value for the infinite bandwidth
Anderson model\cite{Shiba}.

Our results have direct implications for the Mott-Hubbard systems. The
conductivity can be estimated by  converting the $\omega^2$ at zero
temperature to $(\pi T)^2$ at finite temperature
and assuming that the self-energy we
derived at the critical point also applies to the case of a hypercubic
lattice in $d$ dimensions\cite{JarCon}. Using the Kubo formula, we
find a resistivity $\rho(T) = A T^2$ where
$A = - {\pi^{5/2} \hbar  a \over e^2 D  } {\partial^2
\Sigma (i\omega_n)  \over \partial (i\omega_n)^2 }$
giving rise to a finite ratio

\begin{eqnarray}
{A \over \gamma^2} = 8.2 \times 10^{3} a  ~(\Omega m)
\label{aoverg}
\end{eqnarray}
where $a$ is the
hypercubic lattice spacing in units of meter.  For
$Sr_{1-x}La_{x}TiO_3$ with $x=0.9$, which is close to the Mott
transition, Eq. (\ref{aoverg}) yields $A/\gamma^2=4.4\times
10^{-6}\Omega m$ when we use
$a=8.3 \times 10^{-10} m$\cite{Bednorz}.
This result is in good agreement with the measured value $A/\gamma^2=8
\times 10^{-6} \Omega m$\cite{Uchida} .

In summary, we have introduced a projective self-consistent approach
to strongly correlated electron systems, which
allows us to determine the critical properties at the Mott
transition in the half-filled Hubbard model. We find
low energy scaling functions
that can be quantitatively related to experimental results in
transition metal oxides.

We thank V.  Dobrosavljevi\'{c}, A.E. Ruckenstein, and X.Y.
Zhang for useful discussions. This work was supported by the NSF under
grant \#DMR 9224000.

\figure{Schematic plot of the spectral functions of the conduction
electrons and the impurity configurations, illustrating the low
and high energy parts.\label{bath}}

\figure{Imaginary part of the Matsubara Green function versus the
rescaled frequency $\tilde \omega_n$ for system size $N= 6,8,10$ and
non-interacting (semi-circular) density of states. The inset shows  the
single particle spectral function as function of the rescaled
frequency $\tilde \omega$ with a broadening $\delta=.01$ for
$N=10$.\label{ImG}}

\figure{Real part of the Matsubara spin susceptibility as a function
of the rescaled frequency $\tilde \omega_n$ for $N=10$.
\label{spin}}

%\end{narrowtext}

\begin{references}
\bibitem{note2}
Comparing our critical interaction ($U_c\approx3.0$) with that
extrapolated from the
numerical solution of the full model ($U_{c2}\sim 3.0$) implies
that the assumption of dispersionless lower and upper Hubbard bands
and the truncation of the canonical transformation to order $({D\over
U})^2$ introduces a total error of not more than a few percent.
\end{references}
\end{document}